\documentclass[3p,times,procedia]{elsarticle}
\usepackage{nupha_ecrc}
\usepackage{amsmath,amssymb,bm}
\usepackage{color}

\volume{00}

\firstpage{1}

\journalname{Nuclear Physics A}

\runauth{M. Kitazawa, et al.}

\jid{nupha}

\jnltitlelogo{Nuclear Physics A}

\usepackage{amssymb}

\usepackage[figuresright]{rotating}

\begin{document}

\begin{frontmatter}



\dochead{XXVIIIth International Conference on Ultrarelativistic Nucleus-Nucleus Collisions\\ (Quark Matter 2019)}

\title{Critical fluctuations in a dynamically expanding heavy-ion collision}


\author[label3,label4]{Masakiyo Kitazawa}
\ead{kitazawa@phys.sci.osaka-u.ac.jp}
\author[label1]{Gr\'egoire Pihan}
\author[label1]{Nathan Touroux}
\author[label1]{Marcus Bluhm}
\author[label1]{Marlene Nahrgang}

\address[label3]{Department of Physics, Osaka University, Toyonaka, Osaka 560-0043, Japan}
\address[label4]{J-PARC Branch, KEK Theory Center, Institute of Particle and Nuclear Studies, KEK, 203-1, Shirakata, Tokai, Ibaraki, 319-1106, Japan}
\address[label1]{SUBATECH UMR 6457 (IMT Atlantique, Universit\'e de Nantes, IN2P3/CNRS), 4 rue Alfred Kastler, 44307 Nantes, France}

\begin{abstract}

  For the discovery of the QCD critical point it is crucial to develop
  dynamical models of the fluctuations of the net-baryon number that
  can be embedded in simulations of heavy-ion collisions.
  In this proceeding, we study the dynamical formation of the critical
  fluctuations of the net-baryon number near the QCD critical point
  and their survival in the late stages in an expanding system.
  The stochastic diffusion equation with a non-linear free energy
  functional is employed for describing the evolution of
  conserved-charge fluctuations along trajectories
  in the crossover and first-order transition regions 
  near the QCD critical point.

\end{abstract}


\end{frontmatter}


\section{Introduction}
\label{sec:intro}

The phase diagram of QCD in the temperature ($T$) and baryon chemical
potential ($\mu_{\rm B}$) plane is believed to have a QCD critical
point and a first-order phase transition line.
Experimental search for these structures is one of the most
challenging subjects that will be realized by the relativistic
heavy-ion collisions~\cite{Bluhm:2020mpc,Asakawa:2015ybt}.
Active experimental analyses for this purpose are ongoing in the
beam-energy scan program at RHIC~\cite{Adamczyk:2014ipa,Adam:2019xmk}.
Future experimental programs, FAIR, NICA, and J-PARC-HI, will also
contribute to this project.

Fluctuations are important experimental observables in the search
for the phase structure of QCD~\cite{Bluhm:2020mpc,Asakawa:2015ybt}.
In equilibrated media the fluctuations diverge at the second-order phase
transition associated with the divergence of the correlation length.
It is also known that the higher-order cumulants characterizing
non-Gaussianity of fluctuations have a sharper enhancement and
characteristic sign changes near the critical point.
The experimental search for these behaviors in the fluctuation
observables has been very active at RHIC~\cite{Adamczyk:2014ipa}.

In heavy-ion collisions, however, the hot and dense systems created
by the collisions are dynamically expanding.
Therefore, the non-equilibrium dynamics of fluctuations plays a 
crucial role when the fluctuations are used for the study of the
QCD phase diagram~\cite{Sakaida:2017rtj,Nahrgang:2018afz}.
In the present work, we focus on the fluctuations of conserved charges
and study their dynamical evolution within a Bjorken expansion
by employing the stochastic diffusion
equation~\cite{Pihan:prep,Nonaka:prep}.
By solving this equation numerically, we explore the effects of
the non-equilibrium dynamics on the evolution along trajectories
in the crossover~\cite{Pihan:prep} and first-order
transition~\cite{Nonaka:prep} regions near the critical point.

\section{Stochastic diffusion equation}
\label{sec:SDE}

In this study, we focus on the evolution of conserved-charge
fluctuations in dynamically expanding systems.
The long-wavelength behavior of the fluctuations of a
conserved charge $n(z,t)$ is well described by 
the stochastic diffusion equation (SDE)~\cite{Nahrgang:2018afz}
\begin{align}
  \partial_t n(z,t)
  = \Gamma \partial_z^2 \frac{\delta F[n]}{\delta n}
  - \partial_z \xi(z,t),
  \label{eq:SDE}
\end{align}
with the free energy functional $F[n]$ and 
the noise term $\xi(z,t)$ satisfying $\langle \xi(z,t) \rangle=0$ and
\begin{align}
  \langle \xi(z_1,t_1) \xi(z_2,t_2) \rangle
  = 2\Gamma T \delta(z_1-z_2) \delta(t_1-t_2).
\end{align}

By substituting a quadratic form of the free energy functional,
$F[n] = (1/2) \int dz \,m^2 (n(z,t))^2$, into Eq.~(\ref{eq:SDE}),
we obtain the conventional form of the SDE
\begin{align}
  \partial_t n(z,t)
  = D \partial_z^2 n(z,t) - \partial_z \xi(z,t),
  \label{eq:SDEgauss}
\end{align}
with the diffusion coefficient $D=\Gamma m^2$.
In Ref.~\cite{Sakaida:2017rtj}, the evolution of the 
Gaussian fluctuations of $n(z,t)$ has been discussed by solving
Eq.~(\ref{eq:SDEgauss}) analytically. 
In this model, however, all non-Gaussian cumulants
$\langle n^m \rangle_{\rm c}$ with $m\ge3$
vanish in equilibrium and this property is not suitable for 
describing their dynamical evolution.

\begin{figure}
  \centering
  \includegraphics[width=.43\linewidth]{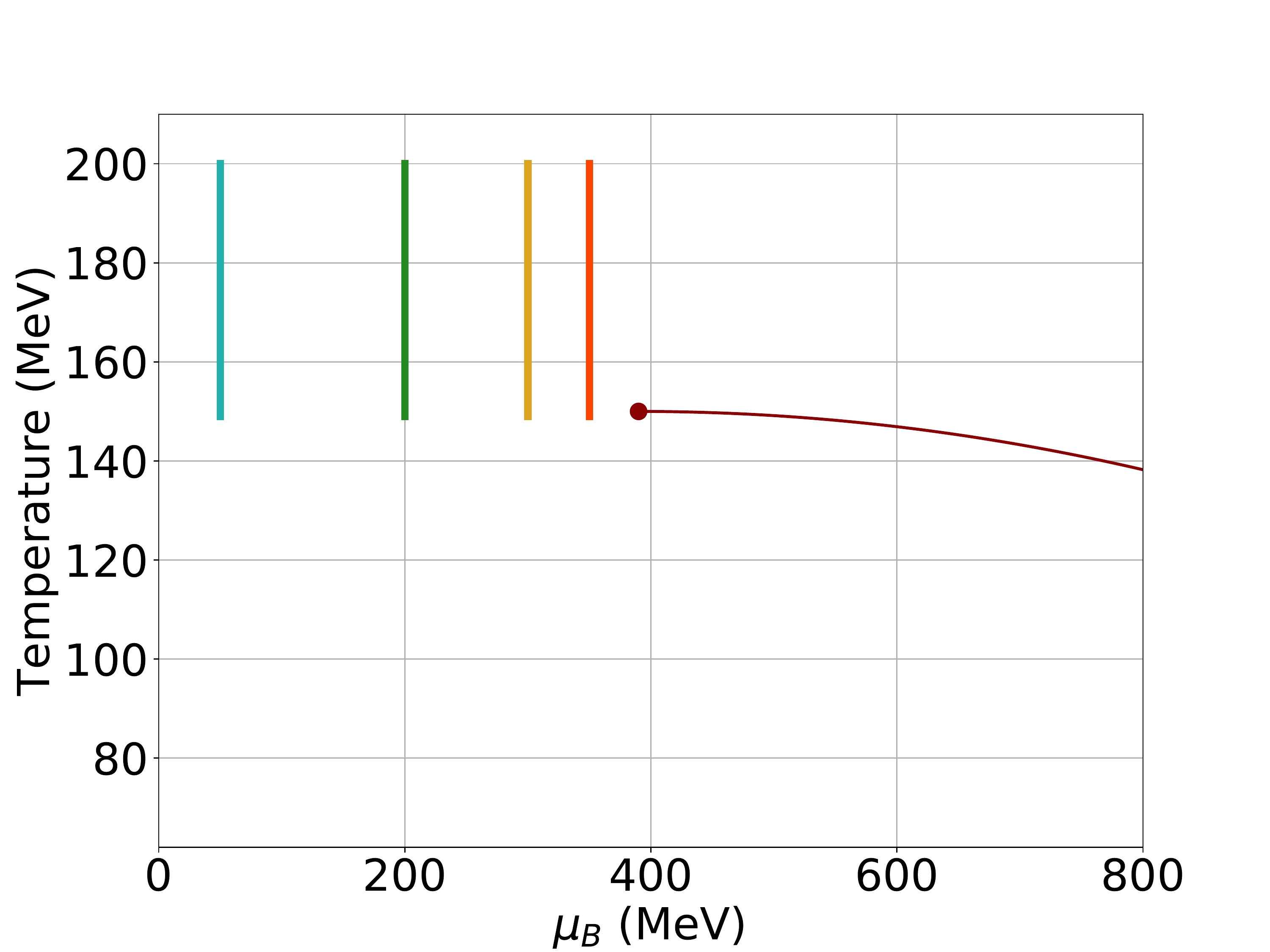}
  \includegraphics[width=.54\linewidth]{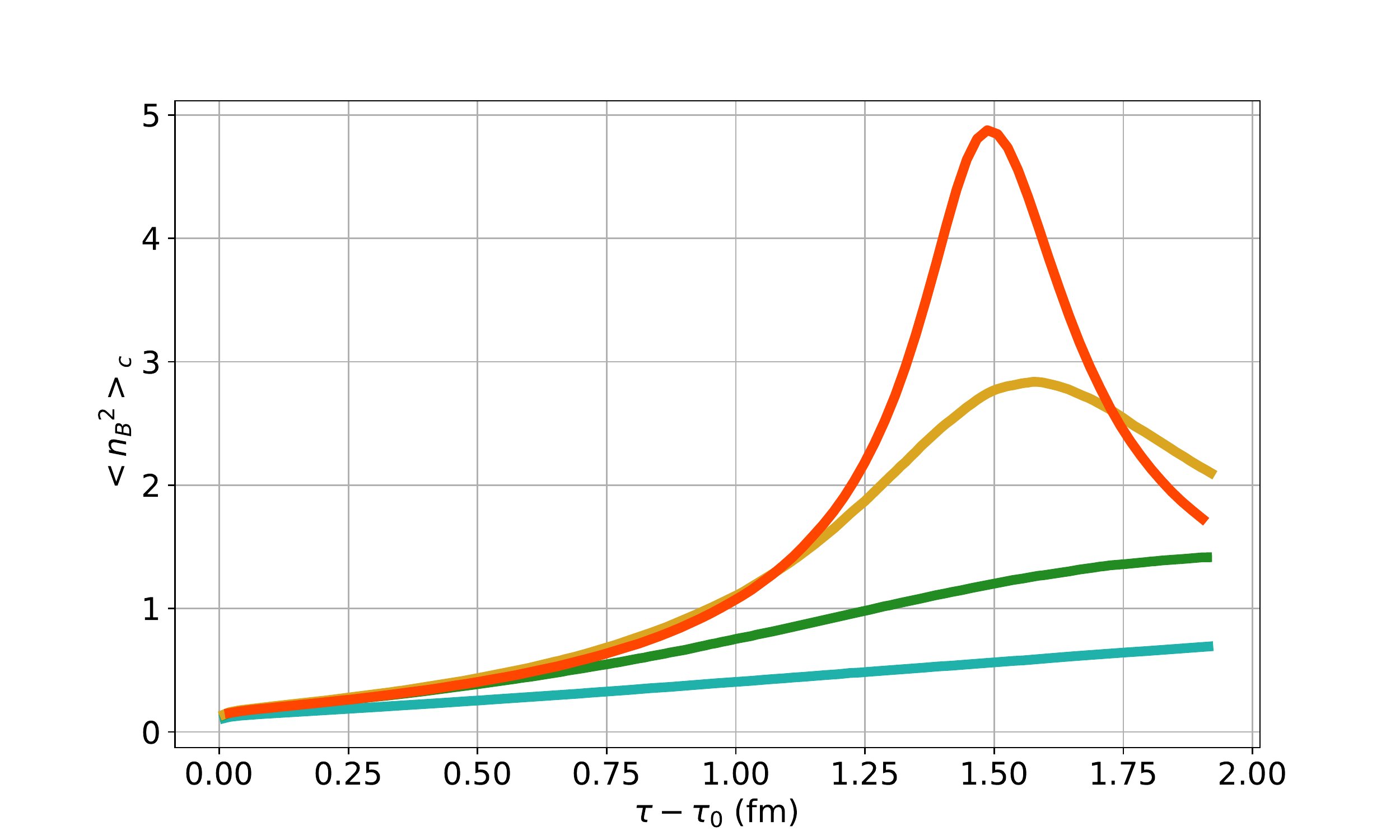}
  \caption{
    Left: Trajectories in the QCD phase diagram discussed in Sec.~\ref{sec:crossover}. 
    The evolutions start at $T_0=200$~MeV and end at $T=149$~MeV. 
    Right: Evolution of the second-order cumulant of the net-baryon number
    $\langle n_{\rm B}^2 \rangle_{\rm c}$ as a function of proper time $\tau$ 
    for simulations without non-linear terms in $F[n]$.
  }
  \label{fig:traj}
\end{figure}

To describe the evolution of non-Gaussian fluctuations,
one has to introduce non-linear terms into $F[n]$.
In Ref.~\cite{Nahrgang:2018afz}, the form of $F[n]$ obtained by
the Taylor expansion around the average density $\langle n \rangle$,
\begin{align}
  F[n] = \int dz \left( \frac{m^2}{2n_c^2} (\delta n)^2
  + \frac{K}{2n_c^2} ( \partial_z \delta n )^2
  + \frac{\lambda_3}{3n_c^3} (\delta n)^3
  + \frac{\lambda_4}{4n_c^4} (\delta n)^4
  + \frac{\lambda_6}{6n_c^6} (\delta n)^6
  \right) ,
  \label{eq:F[n]}
\end{align}
with $\delta n = n - \langle n \rangle$ 
has been employed for describing the non-Gaussianity.
As the SDE is no longer solved analytically with the non-linear terms,
the SDE with Eq.~(\ref{eq:F[n]}) is solved numerically
in the Cartesian coordinate system.
The proper description of the evolution of non-Gaussian fluctuations
has been confirmed~\cite{Nahrgang:2018afz,Nouhou:2019nhe}.

In the following, we consider the evolution of the conserved charge in 
Bjorken-expanding systems assuming boost invariance.
We employ Milne coordinates, i.e.~space-time rapidity $y$ and proper time $\tau$.
In this coordinate system, Eq.~(\ref{eq:SDE}) is rewritten as 
\begin{align}
  \partial_\tau n
  = \frac{\Gamma}{\tau^2} \partial_y^2 \frac{\delta F[n]}{\delta n}
  - \frac1\tau \partial_y \xi -\frac{n}\tau.
  \label{eq:SDEeta}
\end{align}
The last term represents the reduction of the average density due to
the expansion.
The noise correlation is also modified as 
$\langle \xi(y_1,\tau_1) \xi(y_2,\tau_2) \rangle
= 2\tau_1\Gamma T \delta(y_1-y_2) \delta(\tau_1-\tau_2)$.

\section{Crossover region}
\label{sec:crossover}

\begin{figure}
  \centering
  \includegraphics[width=.95\linewidth]{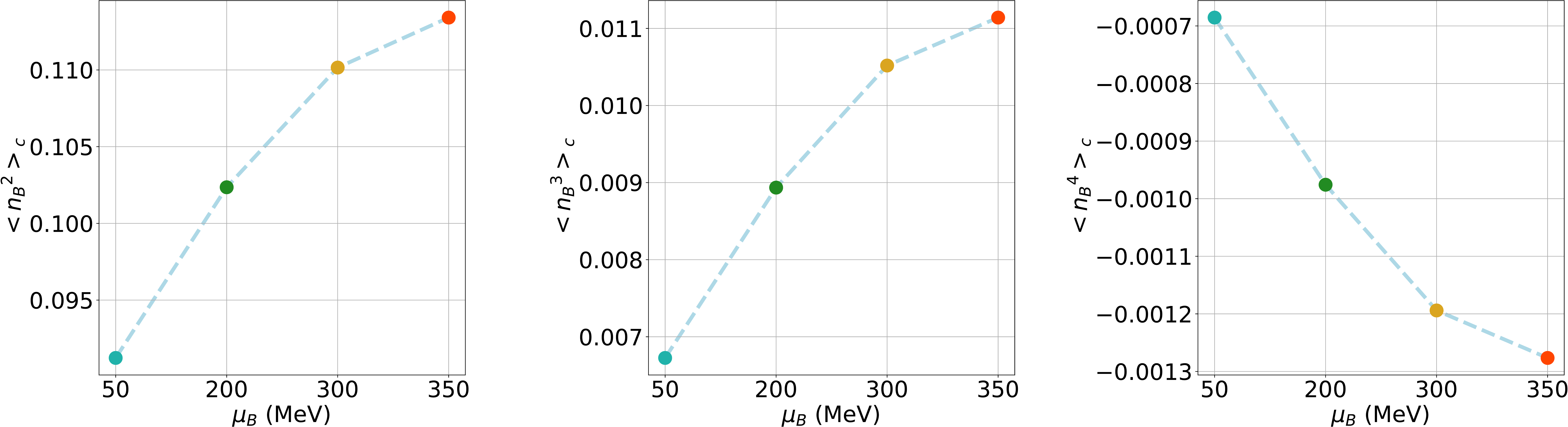}
  \caption{
    Cumulants of the net-baryon number
    $\langle n_{\rm B}^m \rangle_{\rm c}$ at $T=149$~MeV
    obtained for the different trajectories in the left panel of
    Fig.~\ref{fig:traj}.
  }
  \label{fig:cumulant}
\end{figure}

Let us investigate the evolution of the net-baryon number density
$n_{\rm B}(y,\tau)$ along trajectories in the crossover
region by solving Eq.~(\ref{eq:SDEeta}) numerically~\cite{Pihan:prep}.
We consider the four trajectories in the QCD phase diagram shown
in the left panel of Fig.~\ref{fig:traj}.
We employ Eq.~(\ref{eq:F[n]}) for the free energy functional,
where the $T$ and $\mu_{\rm B}$ dependence of the parameters
$\Gamma$, $m^2$, $K$ and $\lambda_i$ has been fixed from the static
universality class of the 3D Ising model and a mapping onto
the QCD phase diagram in line with Ref.~\cite{Nahrgang:2018afz}.
The QCD critical point is located at $T=150$~MeV and $\mu_{\rm B}=390$~MeV.
We set the initial proper time and temperature as $\tau_0=1$~fm and
$T_0=200$~MeV.
The relation between $\tau$ and $T$ is
assumed to be $T(\tau)=T_0(\tau_0/\tau)^{1/3}$.

We first performed numerical simulations without the non-linear terms
and compared the numerical results with the analytic
solution~\cite{Sakaida:2017rtj} to check the correct implementation
of our code.
We verified that the numerical and analytic solutions show
an accurate agreement.
In the right panel of Fig.~\ref{fig:traj}, we show the evolution
of the second-order cumulant $\langle n_{\rm B}^2 \rangle_{\rm c}$
as a function of $\tau$ for the four different trajectories.
In our parametrization of $T(\tau)$, the medium passes through the
transition line at $\tau-\tau_0\simeq1.37$~fm,
while $\langle n_{\rm B}^2 \rangle_{\rm c}$ has a peak
at $\tau-\tau_0\simeq1.5$~fm for the trajectory closest to the critical point. 
This difference highlights a retardation effect.

In Fig.~\ref{fig:cumulant} we show the cumulants
$\langle n_{\rm B}^m \rangle_{\rm c}$
for $m=2$, $3$, and $4$ at $T=149$~MeV just below the transition 
for the four considered trajectories taking the non-linear terms into account.
The figure shows that the absolute values of
$\langle n_{\rm B}^m \rangle_{\rm c}$ are nonzero and become large
as the trajectory approaches the critical point.
These results show that our numerical simulations of 
Eq.~(\ref{eq:SDEeta}) with Eq.~(\ref{eq:F[n]}) reproduce the 
expected critical enhancement of the cumulants.

\section{First-order transition}
\label{sec:1st}

\begin{figure}
  \centering
  \includegraphics[width=.5\linewidth]{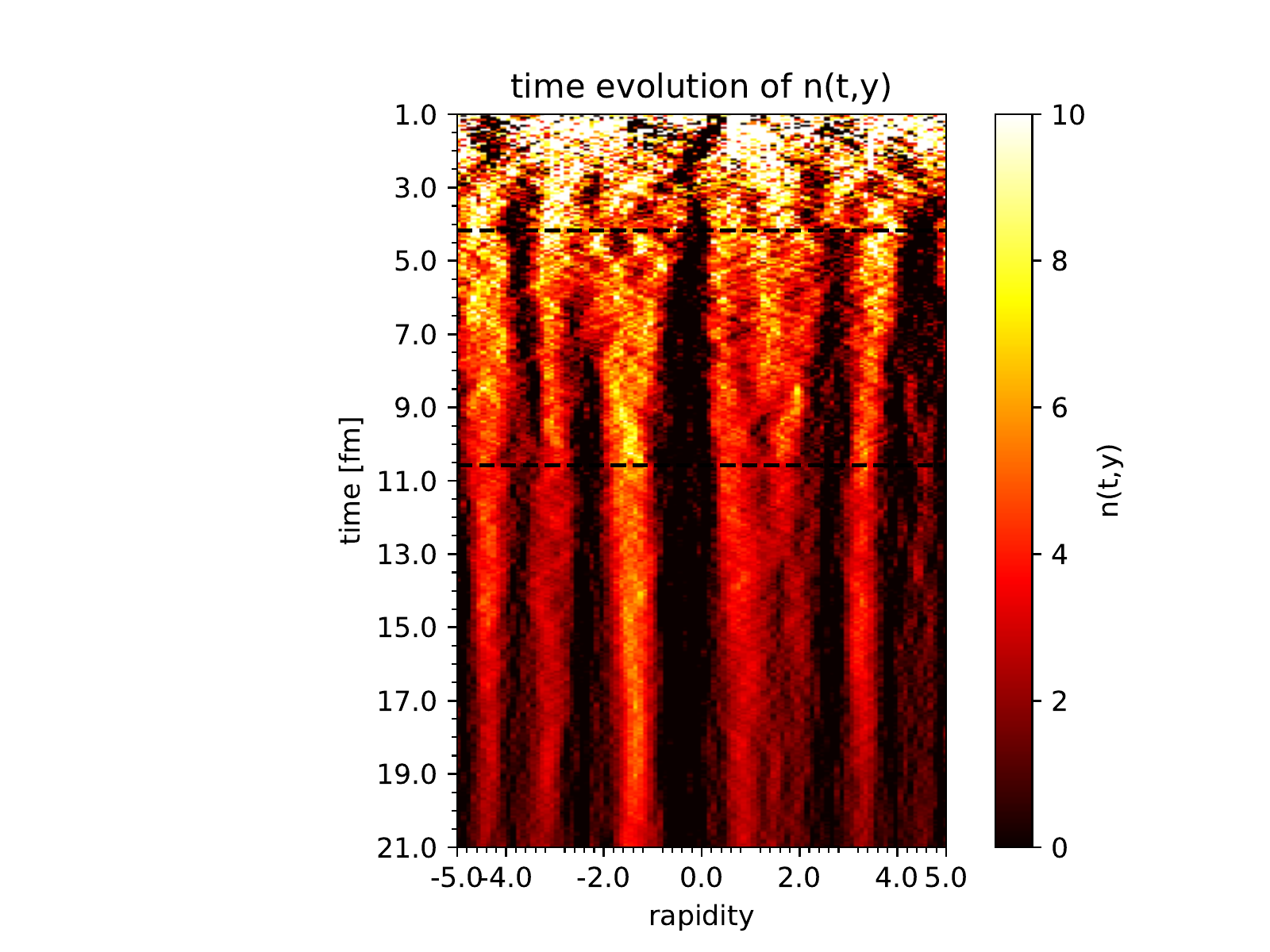}
  \includegraphics[width=.49\linewidth]{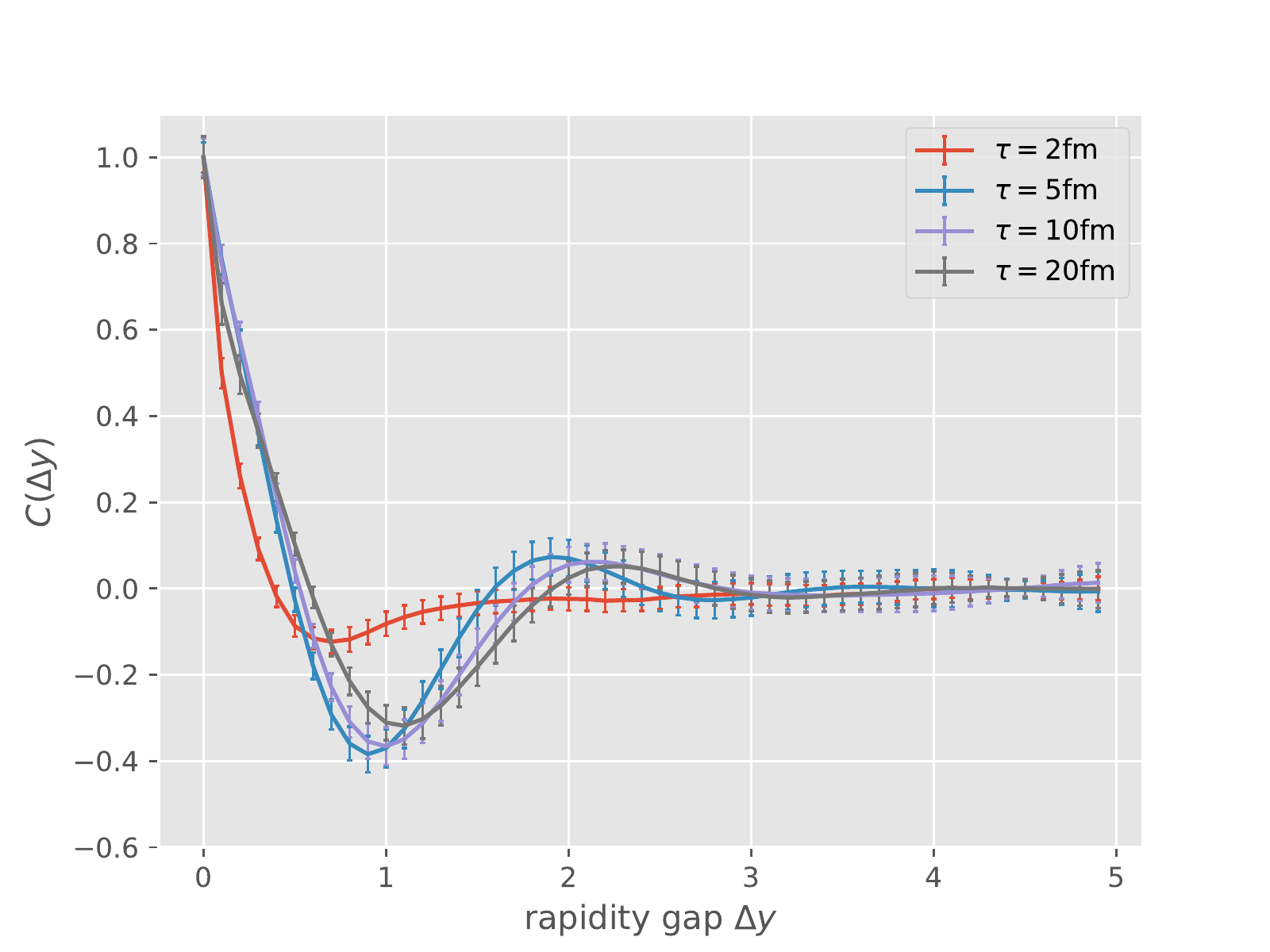}
  \caption{
    Left: Time evolution of the density profile along a trajectory
    across the first-order phase transition. The
    two black horizontal dashed lines show the duration of the first-order
    transition.
    Right: Correlation function $C(\Delta y,\tau)$ at several values of $\tau$.
    }
  \label{fig:1st}
\end{figure}

Let us now study the evolution along a trajectory across the
first-order phase transition~\cite{Nonaka:prep}.
Near the first-order transition $F[n]$ has two local minima
and the global minimum flips at the phase boundary.
To model this behavior of $F[n]$, we employ the following functional form
\begin{align}
  F[n] = \int dz \left( \frac a2 (n-n_s)^2 + \frac b4 (n-n_s)^4
  - c n + K ( \partial_z n )^2 \right) \,,
  \label{eq:F[n]1st}
\end{align}
where $a<0$, $b$, $n_s$, $c$, and $K$ are parameters.
For $c=0$, Eq.~(\ref{eq:F[n]1st}) has two degenerate local minima around
the local maximum at $n=n_s$, while the left (right) local minimum
becomes the global minimum for $c>0$ ($c<0$). 
The coefficient $\Gamma$ is chosen in such a way that
the $T$ dependence of the diffusion coefficient is consistent with
the behavior employed in Ref.~\cite{Sakaida:2017rtj}.

In the left panel of Fig.~\ref{fig:1st} we show an example of the
time evolution of the density profile obtained with Eq.~(\ref{eq:F[n]1st}).
The two black horizontal dashed lines show the proper times at which the average
density is in a local minimum for $c=0$: the medium is in the mixed phase
between these two lines.
The panel shows that the domain formation due to the phase separation
manifests itself with the first-order transition, and the density
inhomogeneity generated by the transition survives even at very large
proper time.

To see effects of the domain formation on observables,
we show the equal-time correlation functions 
$C(\Delta y,\tau) = \langle \delta n(\Delta y,\tau) \delta n(0,\tau) \rangle$
for different $\tau$ in the right panel of Fig.~\ref{fig:1st}.
The panel shows that $C(\Delta y,\tau)$ has a local maximum around
$\Delta y=2.0$ corresponding to the typical size of a domain,
and this structure survives even at the late time $\tau=20$~fm.
This result suggests that such a structure in $C(\Delta y,\tau)$ 
can be used as a signal for the existence of the first-order transition.

This work is in part supported by the TYL-FJPPL joint research program
and JSPS KAKENHI Grant Numbers 17K05442 and 19H05598 as well as 
the program "Etoiles montantes en Pays de la Loire 2017".



\bibliographystyle{elsarticle-num}
\bibliography{ref}







\end{document}